\documentclass[preprint,showpacs,preprintnumbers,amsmath,amssymb]{revtex4}

\usepackage{graphicx}
\usepackage{dcolumn}
\usepackage{bm}

\begin{document}

\draft

\title{Non-Markovian diffusion over a parabolic potential barrier: influence of the friction-memory function}

\author{B. Yilmaz$^{1,2},$ S. Ayik$^{3},$ Y. Abe$^{4},$ and D. Boilley$^{5}$}
\address{$^{1}$Physics Department, Ankara University, 06100 Ankara, Turkey}
\address{$^{2}$Physics Department, Middle East Technical University, 06531 Ankara, Turkey}
\address{$^{3}$Physics Department, Tennessee Technological University, Cookeville, TN 38505, USA}
\address{$^{4}$Research Center for Nuclear Physics, Osaka University, Osaka, Japan}
\address{$^{5}$Grand Acc\'el\'erateur National d'Ions Lourds (GANIL), CEA/DSM-CNRS/IN2P3, BP 55027, F-14076 Caen Cedex 5, France}

\date{\today}

\begin{abstract}
The over-passing probability across an inverted parabolic potential
barrier is investigated according to the classical and quantal
generalized Langevin equations. It is shown that, in the classical
case, the asymptotic value of the over-passing probability is
determined by a single dominant root of the "characteristic
function", and it is given by a simple expression. The expression
for the over-passing probability is quite general, and details of
dissipation mechanism and memory effects enter into the expression
only through the dominant root of the characteristic equation.
\end{abstract}


\maketitle

\section{Introduction}

In many physical systems, for example transport processes in
condensed matter physics, activation processes in chemical
reactions, and thermal fission and fusion reactions in nuclear
physics, generalized Langevin approach provides a very useful
framework for theoretical description of the reaction under
consideration
\cite{Gardiner,Weiss,Abe2,Hanggi2,Grabert,Percival,Caldeira}.
According to general framework of Mori \cite{Mori}, the equations of
motion of relevant variables, in general, appear as non-Markovian
stochastic differential equations, referred to as Generalized
Langevin Equations (GLE). These equations involve memory dependent
dissipation and correlated random forces, which are connected to
each other in accordance with the fluctuation-dissipation relation
of non-equilibrium statistical mechanics. It is possible to derive
the GLE in the classical limit, and also including quantum
statistical effects
\cite{Takigawa,Ayik,Washiyama,Hofmann,Rummel,Grigolini}. The
generalized Langevin approach has recently gained a lot of interest
as a mathematical tool to deal with diffusion in disordered medium,
phenomenon known as anomalous diffusion which is characterized by a
long-range power-law correlations encountered in various physical
processes such as the dynamics of polymers \cite{Amblard},
decorrelation processes in microemulsions \cite{Sciortino}, charge
transport in amorphous semiconductors \cite{Gu}, and diffusion in
fractals \cite{Stephenson}.

After the pioneering work of Kramers, the Langevin approach has been
applied to describe normal as well as anomalous diffusion over a
potential barrier in many research subjects. In order to solve the
GLE, one must assume a particular form for the spectral density of
the environment or the memory kernel which define the non-Markovian
effects. In our study, we investigate the consequences of the
non-Markovian effects on the asymptotic behavior of the system. We
consider the evolution of a single-relevant variable with sharp
initial values according to the classical and quantal GLE. The noise
term in the GLE is a Gaussian stochastic variable and hence the
probability distribution has a Gaussian form, which is specified by
the mean-values and the variances of the relevant variables. In the
specific case of exponential Friction-Memory Function (FMF), the
non-Markovian problem was solved analytically in \cite{Boilley2}.
Here, we consider a general form of the FMF, and investigate memory
effects on the dynamical evolution of normal as well as anomalous
systems. We show that for classical GLE the asymptotic value of the
over-passing probability is determined by a single dominant root of
the "characteristic function" Eq. (\ref{D}) and given by a simple
expression Eq. (\ref{prob3}). This expression for the over-passing
probability is quite general, and details of dissipation mechanism
and memory effects enter into the expression only through the
dominant root of the characteristic equation \cite{Boilley2}. In the
case of quantal GLE, the asymptotic expression of the over-passing
probability has the same structure as the classical case, except it
involves a quantity which is determined by a numerical integration
over the spectral density.

The formal expression for the over-passing probability is derived in
Section II. The analysis of the probability and some general results
are explained in Section III. The conclusion is given in Section IV.

\section{Formal expression for the over-passing probability}
\subsection{The over-passing probability for the classical GLE}
The classical GLE reads
\begin{eqnarray}
\ddot{q}(t)=-\frac{1}{m}\frac{\partial V}{\partial
q}-\int_{0}^{t}\chi(t-t^{\prime })\dot{q}(t^{\prime })dt^{\prime
}+\epsilon (t), \label{GLE}
\end{eqnarray}%
where $\chi(t)$ is the model dependent FMF and the stochastic
driving term $\epsilon(t)$ has a Gaussian distribution with first
and second moments given by
\begin{eqnarray}
\langle \epsilon (t)\rangle &=&0,  \label{noise} \\
\langle \epsilon (t)\epsilon (t^{\prime })\rangle &=&\frac{T}{m}\chi
(|t-t^{\prime }|),  \label{noise2}
\end{eqnarray}%
so that the classical fluctuation-dissipation theorem is satisfied.
Here, $T$ is the temperature of the heat-bath. All throughout the
paper, we set $k_B=1$ where $k_B$ is the Boltzmann constant.

For a quadratic potential barrier with a barrier height $B$ which is
defined by the initial position $q_{0}$ as
\begin{eqnarray}
V(q)=\frac{1}{2}m\Omega ^{2}(q_{0}^{2}-q^{2})=B-\frac{1}{2}m\Omega
^{2}q^{2}\label{potential},
\end{eqnarray}
using the Laplace transform of Eq. (\ref{GLE}), the mean and the
variance of $q(t)$ over the noise, denoted by $\langle..\rangle$,
can be obtained as
\begin{eqnarray}
\langle q(t)\rangle= q_{0}\left[1+\Omega^2\int_0^t h(t')dt'\right]
+\frac{p_0}{m}h(t), \label{mean_q}
\end{eqnarray}%
and
\begin{eqnarray}
\sigma _{qq}(t)=\frac{T}{m}\int_0^t dt'\int_0^t
dt''h(t')h(t'')\chi(|t'-t''|), \label{var_q}
\end{eqnarray}
respectively. The time-dependent function $h(t)$ reads
\begin{eqnarray}
h(t)&=&{\cal L}^{-1}\left[1/D(s)\right]\nonumber \\
&=&\sum_i\text{Res}[\tilde{h}(s_i)] e^{s_it},\label{secular}
\end{eqnarray}
where ${\cal L}^{-1}$ stands for inverse Laplace transform and
$\text{Res}[\tilde{h}(s_i)]$ is the residue of $\tilde{h}(s)={\cal
L}[h(t)]$ at the roots (poles) $s_i$ of the characteristic function
\begin{eqnarray}
D(s)=s^2+s\tilde{\chi}(s)-\Omega^2, \label{D}
\end{eqnarray}
with $\tilde{\chi}(s)={\cal L}[\chi(t)]=\int_0^\infty
\exp{(-st)}\chi(t)dt$ being the Laplace transform of $\chi(t)$. The
initial position $q_0$ as well as the initial momentum $p_0$ of the
collective system are considered to be sharp. The formal expressions
for $\langle p(t)\rangle$, $\sigma_{pp}(t)$ and $\sigma_{qp}(t)$ can
also be obtained, but are irrelevant for finding the over-passing
probability. For a quadratic potential, the relevant variables have
Gaussian distribution. By integrating out the momentum, the reduced
distribution
\begin{eqnarray}
 W(q)=\frac{1}{\sqrt{2\pi\sigma_{qq}(t)}}
 \exp{\left(-\frac{(q-\langle
 q(t)\rangle)^2}{2\sigma_{qq}(t)}\right)}
\label{rGaus}
\end{eqnarray}
is obtained. Starting with the initial value $q_0<0$, the
over-passing probability is simply the probability that the system
is found on the other side of the potential barrier, hence the
probability reads \cite{Abe,Boilley},
\begin{eqnarray}
 P(t)&=&\int_0^\infty W(q)dq\nonumber\\
 &=&\frac{1}{2}\,\textrm{Erfc}\left\{ -\frac{\langle q(t)\rangle }{\sqrt{%
2\sigma _{qq}(t)}}\right\}. \label{prob}
\end{eqnarray}
This converges to a finite value,
\begin{eqnarray}
P=P(t\rightarrow \infty )=\frac{1}{2}%
\,\text{Erfc}\left\{ -\frac{\langle q(t\rightarrow \infty )\rangle }
{\sqrt{2\sigma _{qq}(t\rightarrow \infty )}}%
\right\} ,  \label{prob2}
\end{eqnarray}%
which defines the asymptotic value of the over-passing probability.

In various studies on activated rate processes the characteristic
function Eq. (\ref{D}) appears
\cite{Hanggi2,Grote,Grote2,Carmeli,Dakhnovskii,Pollak,Hanggi,Hanggi3}
and it is shown that Eq. (\ref{D})
has only one positive root (or pole), called hereafter $s_1$, which
is larger than the real parts of all the others, see the appendix of
\cite{Carmeli} for details. This suggests that the asymptotic
behavior of Eq. (\ref{secular}) is
\begin{eqnarray}
h(t\rightarrow \infty)=\text{Res}[\tilde{h}(s_1)]
e^{s_1t}.\label{asy}
\end{eqnarray}
Therefore using Eq. (\ref{asy}), the Eqs. (\ref{mean_q}) and
(\ref{var_q}) read
\begin{eqnarray}
\langle q(t\rightarrow \infty)\rangle=
\text{Res}[\tilde{h}(s_1)]\left(\frac{q_{0}\Omega^{2}}{s_1}
+\frac{p_{0}}{m}\right)e^{s_1t}, \label{asy_mean_q}
\end{eqnarray}%
and
\begin{eqnarray}
\sigma _{qq}(t\rightarrow
\infty)=(\text{Res}[\tilde{h}(s_1)])^2\frac{T}{m}
\left(\frac{\Omega^2-s_1^2}{s_1^2}\right)e^{2s_1t},
\label{asy_var_q}
\end{eqnarray}
where the last equation is obtained by performing the integration of
Eq. (\ref{var_q}) using the variables $u=t'+t''$ and $v=t'-t''$ and
then using the equation $D(s_1)=0$. Substituting Eqs.
(\ref{asy_mean_q}) and (\ref{asy_var_q}) into Eq. (\ref{prob2}) we
get the over-passing probability in the form
\begin{eqnarray}
P=\frac{1}{2}\,\text{Erfc}\left\{ \frac{1}{\sqrt{1-y^{2}}}\left(
\sqrt{\frac{B}{T}}-y\sqrt{\frac{K}{T}}\right) \right\},
\label{prob3}
\end{eqnarray}
where $B$ is the barrier height measured from the initial position
defined in Eq. (\ref{potential}) whereas $K=p_0^2/2m$ is the initial
kinetic energy. The function $y$ is given by
\begin{eqnarray}
y=\frac{s_1}{\Omega}.\label{y}
\end{eqnarray}
If $p_0>0$, there exists a critical initial kinetic energy
$K_c=B/y^2$ for which the mean trajectory Eq. (\ref{asy_mean_q})
converges to the top of the barrier, $\langle q(t\rightarrow
\infty)\rangle=0$ and hence the kernel of the error function in Eq.
(\ref{prob3}) vanishes to give the probability $P=1/2$. Since the
critical kinetic energy must be larger than the barrier height $B$,
the function $y$ can assume any value in the interval,
\begin{eqnarray}
0\leq y\leq 1.\label{condition}
\end{eqnarray}
\begin{figure}[h]
\includegraphics[scale=1.5,clip]{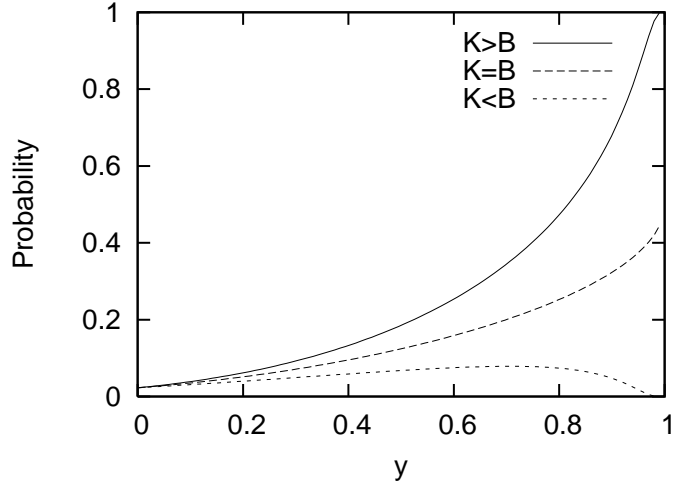}
\caption{The asymptotic probability, Eq. (\protect\ref{prob3}), is
plotted versus the parameter $y$ for the three energy regions. The
temperature is taken such that $T/B = 0.5$ and the solid, dashed and
dotted lines stand for $K/B = 1.5,\,1,\,0.5$ values, respectively.}
\label{proby}
\end{figure}
The function $y$ is the Kramers factor in the rate formula for a
non-Markovian escape process from a metastable state
\cite{Grote,Grote2,Hanggi,Carmeli,Dakhnovskii,Pollak,Hanggi2,Tannor}
and here it is the function that determines the non-Markovian
effects on the over-passing probability. Figure \ref{proby} shows
the probability, Eq. (\ref{prob3}), being plotted versus $y$ for
three energy regions. For $y=1$, the probability takes its classical
value, which correspond to the trivial system without dissipation.
The probability for the over-damped system is obtained for $y=0$,
which is the smallest probability for kinetic energies $K$ larger or
equal to the barrier height $B$. Then, the function $y$ can be
termed as ``the dissipation reducing factor". However, when the
kinetic energy is smaller than the barrier height we have a more
interesting situation where the maximum
\begin{eqnarray}
P_{max}= \frac{1}{2}\,\text{Erfc}\left\{
\sqrt{\frac{B-K}{T}}\right\}\qquad K\leq B, \label{max_prob}
\end{eqnarray}
occurs at some mid-point value
\begin{eqnarray}
y_{max}= \sqrt{\frac{K}{B}}\qquad K\leq B. \label{y_max}
\end{eqnarray}

Since the function $y$ is the only positive root $s_1$ of Eq.
(\ref{D}) divided by the curvature parameter $\Omega$ of the
potential barrier, it depends on the specific form of the FMF and is
a function of the parameters that FMF is expressed by as well as the
curvature parameter $\Omega$ of the potential. There will be a
specific set of these parameters for which $y_{max}$ in Eq.
(\ref{y_max}) will be obtained. The probability $P_{max}$ occurs due
to the compensation between dissipation which reduces probability
and fluctuation which enhances probability. This is explained in the
next section.

It should be emphasized here that the results Eq. (\ref{prob3}) and
Eq. (\ref{max_prob}) are valid for any FMF $\chi(t)$ whose Laplace
transform exists. The formal simplicity of these expressions is due
to the fact that the asymptotic behavior ($t\rightarrow\infty$) of
non-Markovian systems with Gaussian noises diffusing over a
parabolic barrier can be reduced to that of Markovian ones with an
effective friction coefficient,
\begin{eqnarray}
\beta_{\text{eff}}=\tilde{\chi}(s_1), \label{beff}
\end{eqnarray}
which contains all non-Markovian effects. This is easily seen when
the normalized root, using Eq. (\ref{D}), is cast into the formal
form,
\begin{eqnarray}
y=
\sqrt{1+\left(\frac{\beta_{\text{eff}}}{2\Omega}\right)^2}-\frac{\beta_{\text{eff}}}{2\Omega},
\label{y_form}
\end{eqnarray}
which has the same form with the Markovian factor Eq. (\ref{a}).

\subsection{The over-passing probability for the quantum GLE}
For systems with quadratic potentials, the difference between the
classical and c-number quantum GLE is the correlation of the
stochastic force \cite{Gardiner,Weiss,Banerjee}, hence the quantum
GLE has the same form with Eq. (\ref{GLE}) but with a mean-zero
Gaussian noise satisfying the correlation
\begin{eqnarray}
\langle \epsilon (t)\epsilon (t^{\prime })\rangle
=\frac{1}{m}\int_{-\infty}^{\infty}\frac{d\omega}{\pi}T^{\star}(\omega)
\hat{\chi}_{\text{real}}(\omega) e^{-i\omega(t-t^{\prime })},
\label{quancorr}
\end{eqnarray}
instead of Eq. (\ref{noise2}). $T^{\star}$ is the effective
temperature given by
\begin{eqnarray}
T^{\star}(\omega)=\frac{\hbar\omega}{2}
\coth\left(\frac{\hbar\omega}{2T}\right), \label{effT}
\end{eqnarray}
and $\hat{\chi}_{\text{real}}(\omega)$ is the real part of the
Fourier transformed FMF
$\hat{\chi}(\omega)=\int_{-\infty}^{\infty}\chi(t)\exp(i\omega
t)dt$. The effective temperature is the mean energy of a quantum
harmonic oscillator and for high temperatures $\hbar\omega\ll 2T$,
it takes its classical value $T^{\star}\rightarrow T$. The full
quantum limit $T^{\star}\rightarrow\hbar\omega/2$ is obtained at low
temperatures $\hbar\omega\gg 2T$ and represents the zero-point
(vacuum) energy.

Since the noise term does not appear in the expression of the mean
position, the Eqs. (\ref{mean_q}) and (\ref{asy_mean_q}) are valid
for quantum systems as well, whereas the variance of the position
takes the form
\begin{eqnarray}
\sigma_{qq}(t)&=&\frac{1}{m}\int_0^t dt'\int_0^t
dt''h(t')h(t'')\nonumber\\
&&\times\int_{-\infty}^{\infty}\frac{d\omega}{\pi}T^{\star}(\omega)
\hat{\chi}_{\text{real}}(\omega) e^{-i\omega(t'-t'')}.
\label{quan_varq}
\end{eqnarray}
By using Eq. (\ref{asy}), the asymptotic value of the variance reads
\begin{eqnarray}
\sigma _{qq}(t\rightarrow
\infty)&=&\frac{1}{m}(\text{Res}[\tilde{h}(s_1)])^2e^{2s_1t}\nonumber\\
&&\times\int_{-\infty}^{\infty}\frac{d\omega}{\pi}T^{\star}(\omega)
\frac{\hat{\chi}_{\text{real}}(\omega)}{\omega^2+s_1^2}.
\label{quan_asyq}
\end{eqnarray}
Substituting Eq. (\ref{asy_mean_q}) and Eq. (\ref{quan_asyq}) into
Eq. (\ref{prob2}), we get the over-passing probability as
\begin{eqnarray}
P=\frac{1}{2}\,\text{Erfc}\left\{ \frac{1}{\sqrt{G(y)}}\left(
\sqrt{B}-y\sqrt{K}\right) \right\}, \label{quanprob}
\end{eqnarray}
where
\begin{eqnarray}
G(y)=y^2\int_{-\infty}^{\infty}\frac{d\omega}{\pi}T^{\star}(\omega)
\frac{\hat{\chi}_{\text{real}}(\omega)}{\omega^2+s_1^2}.
\label{Gfun}
\end{eqnarray}

\section{Analysis of the Probability}

\subsection{Influence of the memory}
The knowledge of the FMF $\chi(t)$ is crucial for determining the
over-passing probability since the probabilities Eq. (\ref{prob3})
and Eq. (\ref{quanprob}) are functions of the positive root of the
characteristic function Eq. (\ref{D}) which depends on the Laplace
transform $\tilde{\chi}(s)$. As an example, in the Markovian (M)
limit, the normalized root $y$ is explicitly given by
\begin{eqnarray}
y^{(\text{M})} =\sqrt{1+\left( \frac{\beta }{%
2\Omega }\right) ^{2}}-\frac{\beta }{2\Omega } \label{a}
\end{eqnarray}
for the FMF \cite{Abe},
\begin{eqnarray}
\chi^{(\text{M})} (t)=2\beta\delta(t),\label{Dirac}
\end{eqnarray}
which corresponds to a memoryless dissipation with a reduced
friction coefficient $\beta$. For non-Markovian Exponential (E) FMF
\cite{Boilley2},
\begin{eqnarray}
\chi^{(\text{E})}(t)= \frac{\beta}{\tau}\exp {\left( -\frac{t}{\tau}
\right) }, \label{exp1}
\end{eqnarray}
the normalized root can be expressed as
\begin{widetext}
\begin{eqnarray}
y^{(\text{E})} &=&-\frac{1}{3\Omega
\tau }  \nonumber \\
&&-\frac{1}{3\Omega \tau }\,\frac{-1+3\beta \tau -3{\left( \Omega
\tau
\right) }^{2}}{{\left[ -1+\frac{9}{2}\beta \tau +9{(\Omega \tau )}^{2}+\sqrt{%
{\left( -1+\frac{9}{2}\beta \tau +9{(\Omega \tau )}^{2}\right)
}^{2}+{\left(
-1+3\beta \tau -3{\left( \Omega \tau \right) }^{2}\right) }^{3}}\right] }%
^{1/3}}  \nonumber \\
&&+\frac{1}{3\Omega \tau }{\left[ -1+\frac{9}{2}\beta \tau +9{(\Omega \tau )}%
^{2}+\sqrt{{\left( -1+\frac{9}{2}\beta \tau +9{(\Omega \tau )}^{2}\right) }%
^{2}+{\left( -1+3\beta \tau -3{\left( \Omega \tau \right) }^{2}\right) }^{3}}%
\right] }^{1/3}.  \label{b}
\end{eqnarray}
\end{widetext}
The exponential FMF corresponds to a dissipating system with a
reduced friction coefficient $\beta$ and a memory characterized by
the relaxation time $\tau$. In the limit $\tau\rightarrow 0^+$ the
exponential FMF Eq. (\ref{exp1}) reduces to the Markovian FMF Eq.
(\ref{Dirac}).

\begin{figure}[htb]
\includegraphics[scale=0.9,clip]{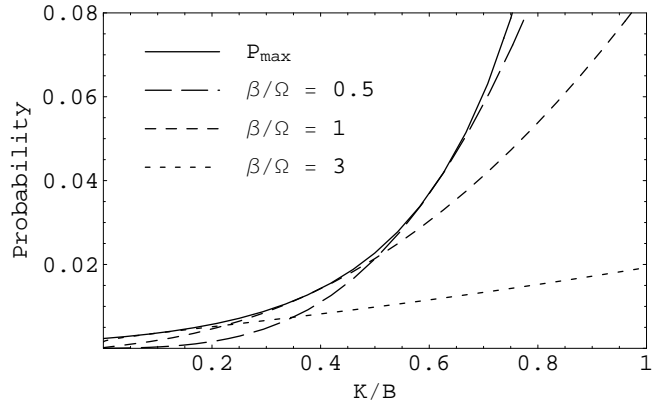}
\caption{The probability for classical Markovian case is plotted
versus the initial kinetic energy over potential barrier height for
various friction coefficients. The maximum probability, Eq.
(\protect\ref{max_prob}), is indicated by a solid line. The
temperature is such that $T/B=0.25$.} \label{probMK}
\end{figure}

\begin{figure}[htb]
\includegraphics[scale=1,clip]{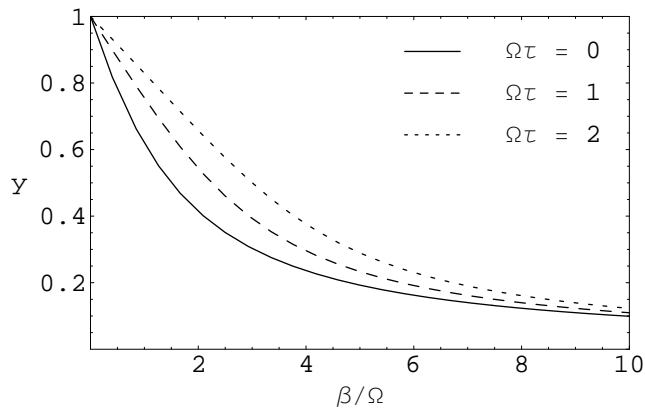}
\caption{The functions Eq. (\ref{a}) (solid line) and Eq. (\ref{b})
are plotted versus the friction $\beta/\Omega$ for different
relaxation times $\Omega\tau$.} \label{yb}
\end{figure}

\begin{figure*}[htb]
\includegraphics[width=8.0cm,clip]{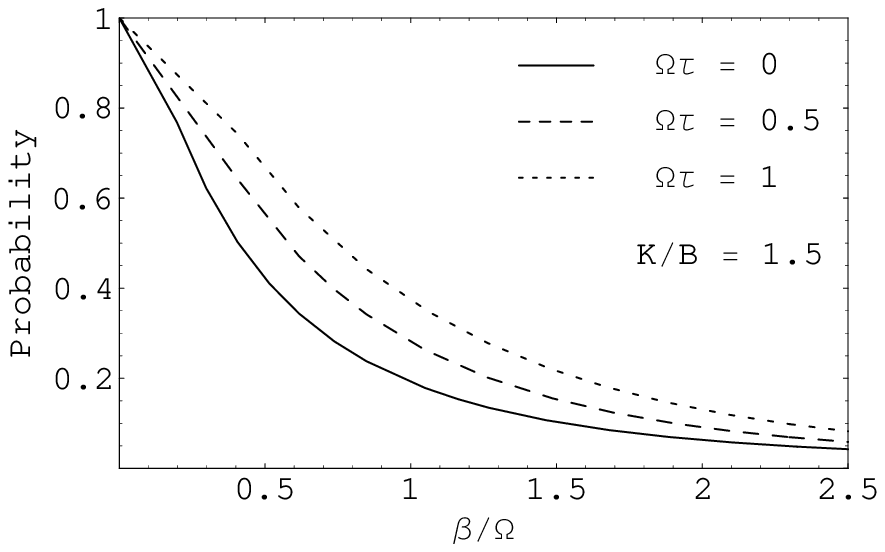} %
\includegraphics[width=8.0cm,clip]{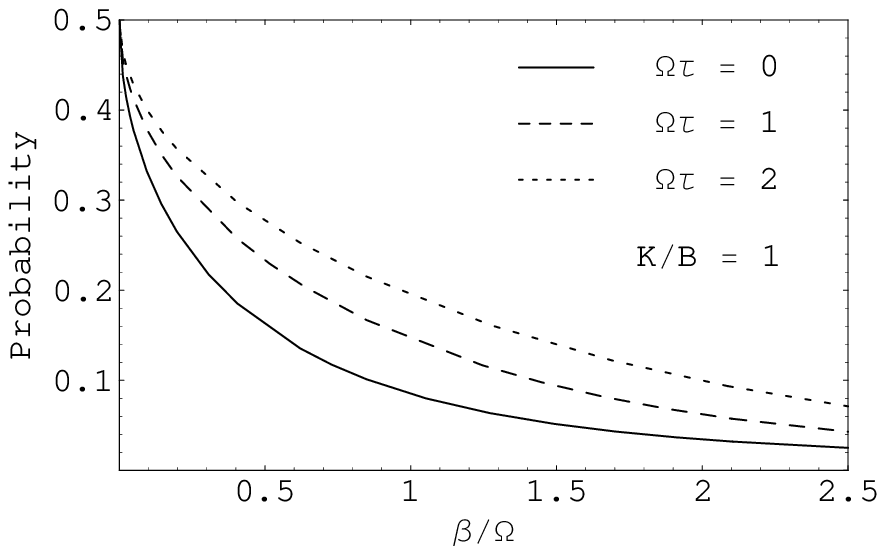} %
\includegraphics[width=8.0cm,clip]{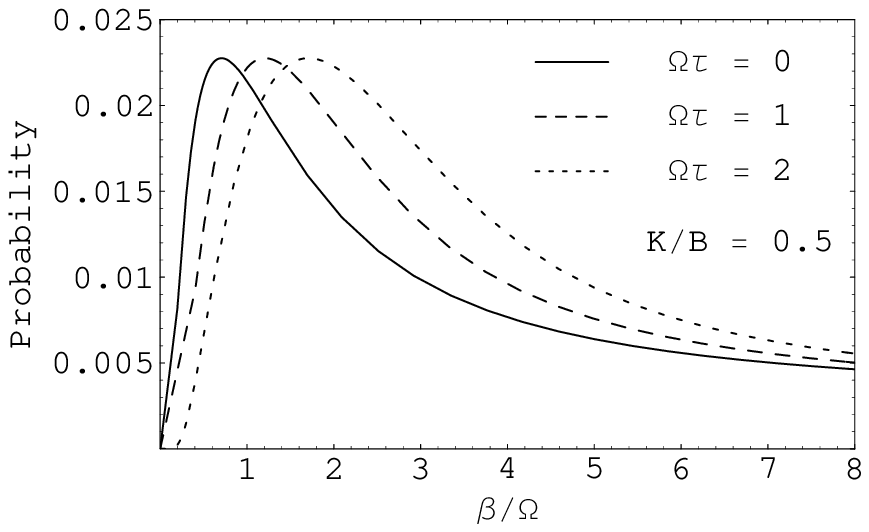} %
\includegraphics[width=8.0cm,clip]{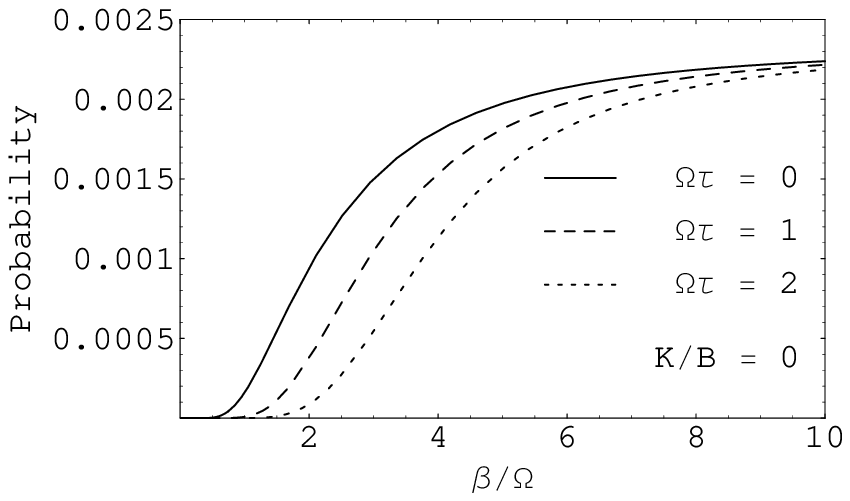}
\caption{The probability Eq. (\ref{prob3}) for the exponential FMF
Eq. (\ref{exp1}) with the corresponding normalized positive root Eq.
(\ref{b}) is plotted versus the friction $\beta /\Omega$ for various
memory times $\Omega\tau$. No memory case $\Omega\tau=0$ (solid
lines) corresponds to the Markovian FMF Eq. (\ref{Dirac}) with $y$
given by Eq. (\ref{a}). The temperature is taken so that $T/B=0.25$
and each figure is plotted with different initial kinetic energy
$K/B=1.5,\,1,\,0.5,\,0$ as labeled on them.} \label{expbo}
\end{figure*}

The over-passing probability for classical systems with Markovian
FMF Eq. (\ref{Dirac}) and the exponential FMF Eq. (\ref{exp1}) is
obtained by substituting Eq. (\ref{a}) and Eq. (\ref{b}) into Eq.
(\ref{prob3}), respectively. Figure \ref{probMK} shows the
probability for the Markovian system plotted versus the initial
kinetic energy over potential barrier height for various friction
coefficients. The intersection points with the maximum probability
(solid line) corresponds to the specific initial kinetic energy and
specific friction $\beta/\Omega$ in Eq. (\ref{a}) for which the
condition Eq. (\ref{y_max}) is met.
In Figure \ref{yb}, it is seen that as friction $\beta/\Omega$
increases, the factors $y^{(\text{M})}$ and $y^{(\text{E})}$ are
decreasing. Increasing memory time $\Omega\tau$ results in
increasing $y$ for any $\beta/\Omega$. Therefore by looking at the
behavior of the probability with respect to the factor $y$ (see
Figure \ref{proby}), the friction $\beta/\Omega$ and memory time
$\Omega\tau$ dependence of the probability is expected as in Figure
\ref{expbo}. For kinetic energies $K$ larger or equal to the barrier
height $B$, the probability is decreasing as $\beta/\Omega$ is
increasing which is due to the dissipation of kinetic energy. For
zero kinetic energy the situation is opposite, as $\beta/\Omega$ is
getting larger the probability is increasing which is due to the
thermal fluctuations of the observables. In the intermediate region
where the kinetic energy is less than the barrier height $B$, the
probability obtains a peaked value, Eq. (\ref{max_prob}). This can
be understood as the dissipation dominating at the right side of the
maximum and fluctuation dominating at the left side of the maximum.
In this region, the probability approaches the asymptotic value
$0.5\,\text{Erfc}\{\sqrt{B/T}\}$ for the over-damped case
$\beta/\Omega\rightarrow\infty$ regardless of the initial kinetic
energy $K$ and memory time $\Omega\tau$.

In general, by using Eq. (\ref{D}) a relation between the root $y$
for any non-Markovian FMF $\chi(t)$ and the Markovian root
$y^{(\text{M})}$ given by Eq. (\ref{a}) can be obtained as
\begin{eqnarray}
\frac{y}{y^{(\text{M})}}=\frac{y^{(\text{M})}+\beta^{(0)}/\Omega}{y
+\beta_{\text{eff}}/\Omega}. \label{condition2}
\end{eqnarray}
Here, the zero-frequency component of the Laplace transformed FMF
defines the static friction,
\begin{eqnarray}
\beta^{(0)}=\tilde{\chi}(0)=\int_0^\infty \chi(t)\,dt\label{marschi}
\end{eqnarray}
which is equal to the friction coefficient $\beta$ in case of normal
(non-anomalous) diffusion and $\beta_{\text{eff}}$ is the
non-Markovian effective friction defined in Eq.(\ref{beff}). From
Eq. (\ref{condition2}) and Eq. (\ref{marschi}), when $y$ is larger
(smaller) than $y^{(\text{M})}$, the non-Markovian effective
friction $\beta_{\text{eff}}$ is smaller (larger) than the Markovian
friction $\beta$ \cite{Grote}. Hence by comparing the non-Markovian
roots with Markovian roots, it is possible to relate the effective
frictions and hence the probabilities.

\subsection{Influence of the oscillations} 
\begin{figure}[htb]
\includegraphics[scale=1,clip]{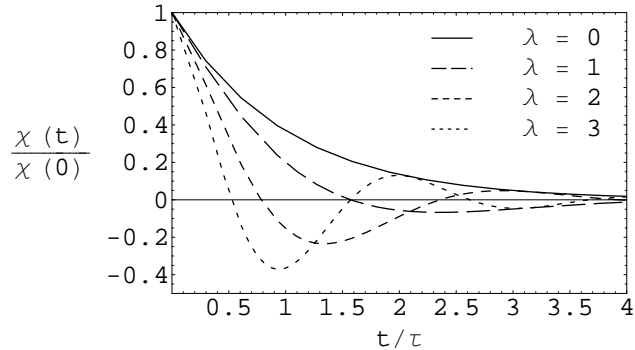}
\caption{The normalized FMF Eq. (\ref{expcos_mem2}) is plotted
versus time. Different lines correspond to different oscillation
frequencies $\lambda$, the solid line with $\lambda=0$ is the
exponential FMF.} \label{expcospl}
\end{figure}

\begin{figure}[htb]
\includegraphics[scale=1,clip]{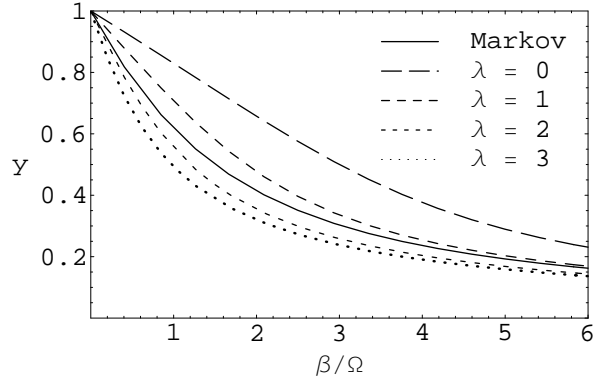}
\caption{The normalized positive root $y^{(\text{EC})}$ of the Eq.
(\ref{D}) where the Laplace transform of Eq. (\ref{expcos_mem2}) is
used, is plotted versus friction $\beta/\Omega$ for various
oscillation frequencies $\lambda$. The memory time is chosen as
$\Omega\tau=2$. The solid line corresponds to the Markovian factor
Eq. (\ref{a}).} \label{yexpcos}
\end{figure}

\begin{figure*}[htb]
\includegraphics[width=8.0cm,clip]{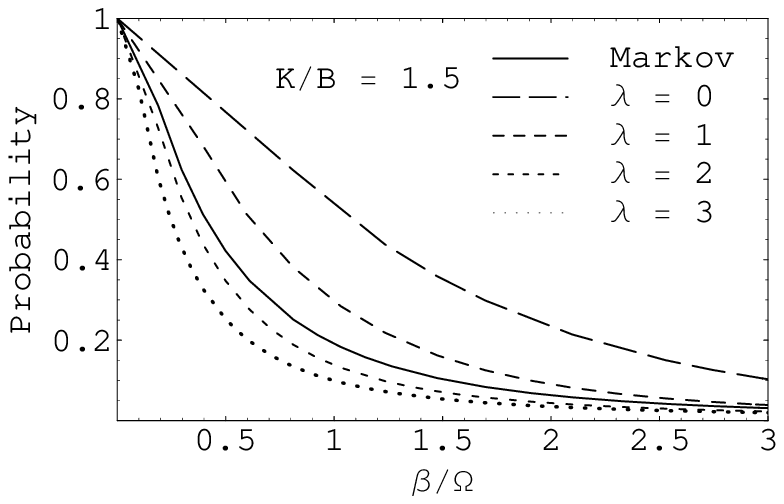} %
\includegraphics[width=8.0cm,clip]{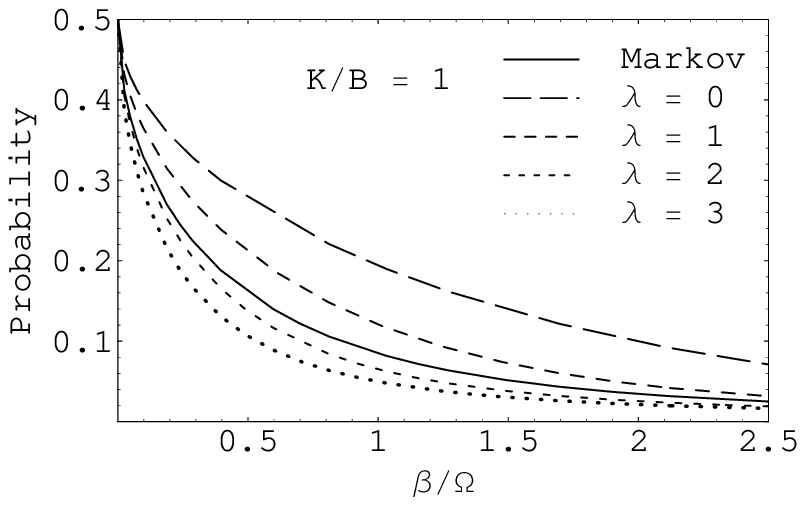} %
\includegraphics[width=8.0cm,clip]{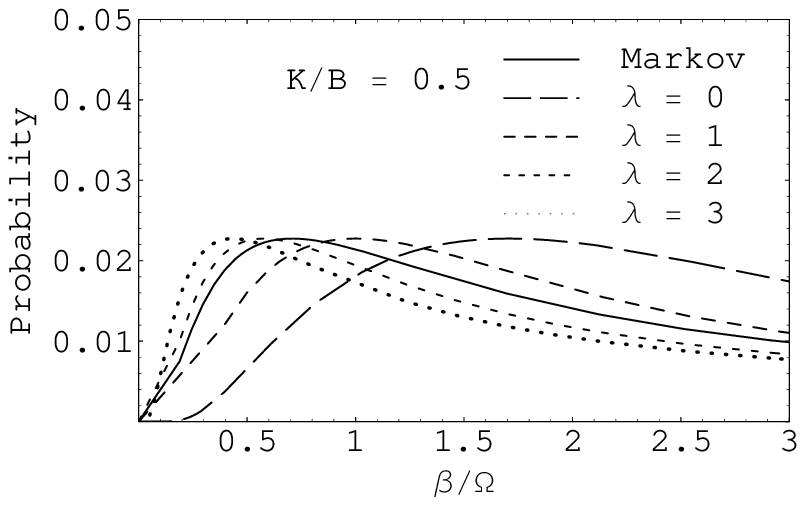} %
\includegraphics[width=8.0cm,clip]{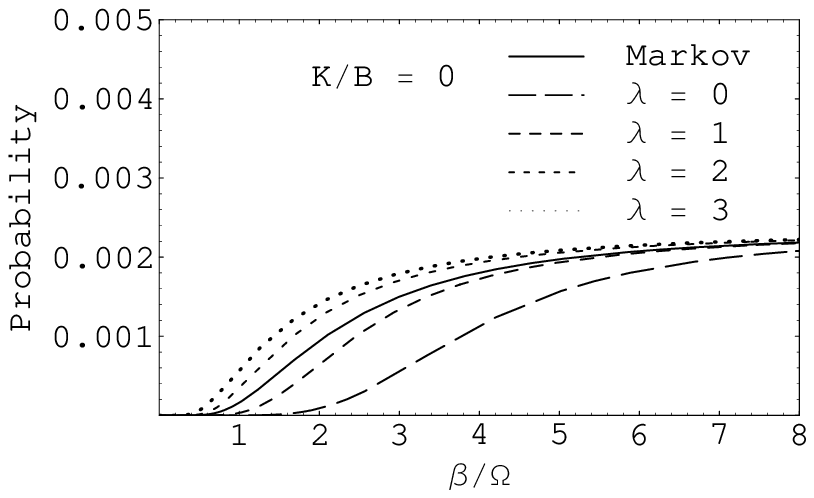}
\caption{The probability Eq. (\ref{prob3}) for the
exponential-cosine FMF Eq. (\ref{expcos_mem2}) with the
corresponding factor $y^{(\text{EC})}$ shown in Figure
\ref{yexpcos}, is plotted versus the friction $\beta /\Omega$ for
various oscillation frequencies $\lambda$. The memory time is chosen
as $\Omega\tau=2$. The Markovian probability is indicated by a solid
line. The temperature is taken so that $T/B=0.25$ and each figure is
plotted with different initial kinetic energy
$K/B=1.5,\,1,\,0.5,\,0$ as labeled on them.} \label{probexpcos}
\end{figure*}

It is possible to obtain
an exact FMF for a system coupled to a heat-bath of harmonic
oscillators known as Caldeira-Leggett Model
\cite{Caldeira,Senitzky,Zwanzig,Ford}. By using this model, the
global degrees of freedom are reduced to the relevant ones and a GLE
in the form of Eq. (\ref{GLE}) is obtained. The corresponding FMF is
a sum of cosine functions and hence has an oscillatory behavior. In
order to understand the consequences of the oscillatory memories,
let us consider the following FMF,
\begin{eqnarray}
\chi^{(\text{EC})}(t)= \frac{(1+\lambda^2)\beta}{\tau}\exp{\left(
-\frac{t}{\tau}\right)}\cos\left(\lambda\frac{t}{\tau}\right),
\label{expcos_mem2}
\end{eqnarray}
where the parameter $\lambda$ keeps track of the oscillations (see
Figure \ref{expcospl}).
The plots of the corresponding normalized positive root
$y^{(\text{EC})}$ where EC stands for Exponential-Cosine FMF Eq.
(\ref{expcos_mem2}) and the corresponding over-passing probability
for classical systems are shown in Figure \ref{yexpcos} and Figure
\ref{probexpcos}, respectively. It is seen that for small values of
$\lambda$ the factor $y^{(\text{EC})}$ is larger than
$y^{(\text{M})}$ (solid line) and for large values of $\lambda$ the
factor $y^{(\text{EC})}$ is smaller than $y^{(\text{M})}$. By using
the equation
$\tilde{\chi}^{(\text{EC})}(s_1)=\tilde{\chi}^{(\text{EC})}(0)$ and
the Laplace transform of Eq. (\ref{expcos_mem2}), the critical value
$\lambda_c$ for which $y^{(\text{EC})}=y^{(\text{M})}$ is satisfied
can be found as
\begin{eqnarray}
\lambda_c=\sqrt{1+\Omega\tau y^{(\text{M})}},\label{lambda}
\end{eqnarray}
which is in the interval $1<\lambda_c<\sqrt{1+\Omega\tau}$. When the
dimensionless oscillation frequency is less than the critical value
$\lambda<\lambda_c$, one has $y^{(\text{EC})}>y^{(\text{M})}$ and
hence non-Markovian dissipation is smaller than the Markovian
dissipation, $\beta_{\text{eff}}<\beta$. This means that the
oscillations of the FMF are irrelevant during the memory time
$\Omega\tau$ and the dissipation is reduced like in the exponential
FMF case. When the oscillations are relevant $\lambda>\lambda_c$,
one has $y^{(\text{EC})}<y^{(\text{M})}$ and consequently the
oscillations of the FMF cancel out in the characteristic time
interval $\Omega\tau$ increasing the non-Markovian effective
friction $\tilde{\chi}^{(\text{EC})}(s_1)$ with respect to the
Markovian friction $\tilde{\chi}^{(\text{EC})}(0)$.

For the asymptotic value $\lambda\rightarrow\infty$, substituting
the Laplace transform of Eq. (\ref{expcos_mem2}) into Eq. (\ref{D})
and taking the limit, the normalized positive root can be found as
\begin{eqnarray}
y^{(\text{EC})}_{\lambda\rightarrow
\infty}=\frac{1}{1+\beta\tau}\left[\sqrt{(1+\beta\tau)+\left(
\frac{\beta }{ 2\Omega }\right) ^{2}}-\frac{\beta }{2\Omega
}\right], \label{a_lambda}
\end{eqnarray}
which satisfies the following inequality $
y^{(\text{EC})}_{\lambda\rightarrow\infty}<y^{(\text{EC})}_{\lambda>\lambda_c}
<y^{(\text{M})}<y^{(\text{EC})}_{\lambda<\lambda_c}$. The
corresponding probabilities follow the same order, $
P^{(\text{EC})}_{\lambda\rightarrow\infty}<P^{(\text{EC})}_{\lambda>\lambda_c}
<P^{(\text{M})}<P^{(\text{EC})}_{\lambda<\lambda_c}$ for $K\geq B$.

\subsection{Influence of the anomalous diffusion}
The FMF for a system coupled to a Non-Ohmic (NO) heat-bath can be
expressed as
\begin{eqnarray}
\chi^{(\text{NO})}(t)&=&2\int_0^\infty
\frac{d\omega}{\pi}\frac{J(\omega)}{\omega}\cos(\omega t),
\label{chigen}
\end{eqnarray}
where $J(\omega)$ is the spectral density of the heat bath.
Non-Ohmic spectral density has the form
\cite{Weiss,Grabert,Grabert2,Leggett,Lutz,Bao,Bao2,Vinales}
\begin{eqnarray}
J(\omega)=\beta\,\frac{\omega^\alpha}{\omega_r^{\alpha-1}}\qquad(0<\alpha<2),
\label{nonohm}
\end{eqnarray}
where $\omega_r$ is some reference frequency allowing for consistent
dimensionality of the friction $\beta$ for any $\alpha$. For matter
of convenience, we set this frequency as that of the potential
barrier, $\omega_r=\Omega$. The explicit form of the FMF is,
\begin{eqnarray}
\chi^{(\text{NO})}(t)=\frac{2\beta}{\pi\Omega^{\alpha-1}}\cos(\frac{\pi\alpha}{2})\,\Gamma(\alpha)\,t^{-\alpha}\qquad
(t\neq 0), \label{nonohmcor}
\end{eqnarray}
with the Laplace transform,
\begin{eqnarray}
\tilde{\chi}^{(\text{NO})}(s)=\frac{\beta}{\sin(\frac{\pi
\alpha}{2})} \left(\frac{s}{\Omega}\right)^{\alpha-1}.
\label{lapnocutoff}
\end{eqnarray}
\begin{figure}[htb]
\includegraphics[scale=1,clip]{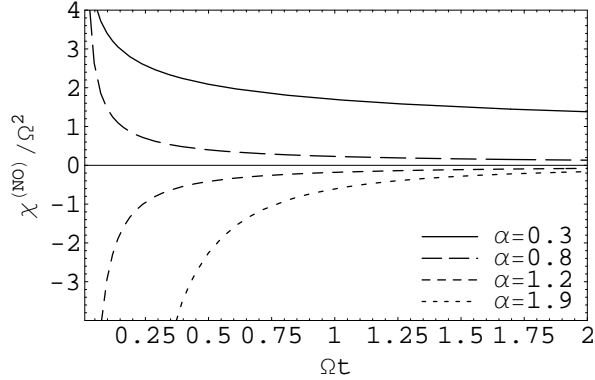}
\caption{The FMF Eq.(\ref{nonohmcor}) is plotted versus time $\Omega
t$ for different values of $\alpha$. The friction coefficient is
taken as $\beta/\Omega=1$.} \label{figa1}
\end{figure}
\begin{figure}[htb]
\includegraphics[scale=0.8,clip]{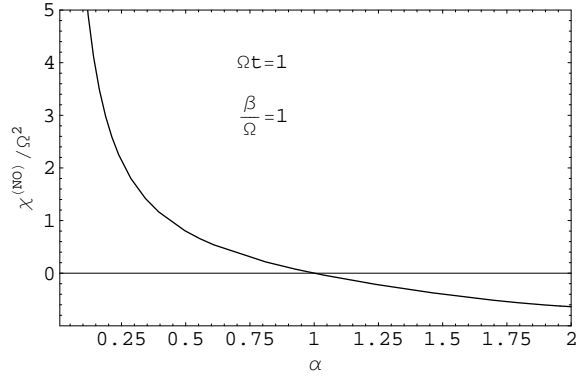}
\caption{The FMF Eq.(\ref{nonohmcor}) is plotted versus the
parameter $\alpha$. The time and friction coefficient are taken as
$\Omega t=1$ and $\beta/\Omega=1$, respectively.} \label{figa2}
\end{figure}
Figures \ref{figa1} and \ref{figa2} show the plot of the FMF Eq.
(\ref{nonohmcor}) as a function of time and $\alpha$, respectively.
Note that $\alpha=1$ recovers the normal Markovian FMF. For
super-Ohmic case $\alpha>1$, the FMF Eq. (\ref{nonohmcor}) is
negative and approaches $-\infty$ as $t\rightarrow 0$, but from Eq.
(\ref{chigen}) it is seen that
$\chi^{(\text{NO})}(0)\rightarrow+\infty$. Furthermore, these
divergences in the super-Ohmic case are such that the static
friction is vanishing,
$\beta^{(0)}=\tilde{\chi}^{(\text{NO})}(0)=\int_0^{\infty}\chi^{(\text{NO})}(t)dt=0$.
On the other hand, for sub-Ohmic diffusion $\alpha<1$ there is a
divergence to $+\infty$ as $t\rightarrow 0$, hence the static
friction is divergent, $\beta^{(0)}=+\infty$. This behavior is
completely different from the Ohmic dissipative systems which have
static frictions that are simply equal to the friction coefficient
$\beta$.

A force-free system coupled to a bath with non-Ohmic spectral
density of the form Eq. (\ref{nonohm}) exhibits anomalous diffusion
\cite{Morgado} which is characterized by the mean square
displacement given by
\begin{eqnarray}
\langle x^2(t)\rangle\sim t^\alpha
\qquad(t\rightarrow\infty),\label{anovarq}
\end{eqnarray}
where, for $0<\alpha<1$ the system is called subdiffusive and for
$1<\alpha<2$ the system is superdiffusive. The static friction can
be used to distinguish between the sub-Ohmic
($\beta^{(0)}\rightarrow\infty$), Ohmic
($\beta^{(0)}\rightarrow\text{finite}$), and super-Ohmic
($\beta^{(0)}\rightarrow 0$) environments which mean sub-diffusion,
normal diffusion, super-diffusion for force-free systems.
\begin{figure}[htb]
\includegraphics[scale=1.3,clip]{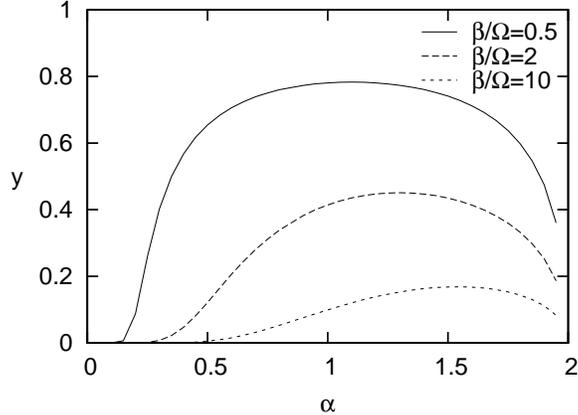}
\caption{The normalized root is plotted versus the parameter
$\alpha$.} \label{figb2}
\end{figure}

\begin{figure*}[htb]
\includegraphics[width=7.5cm,clip]{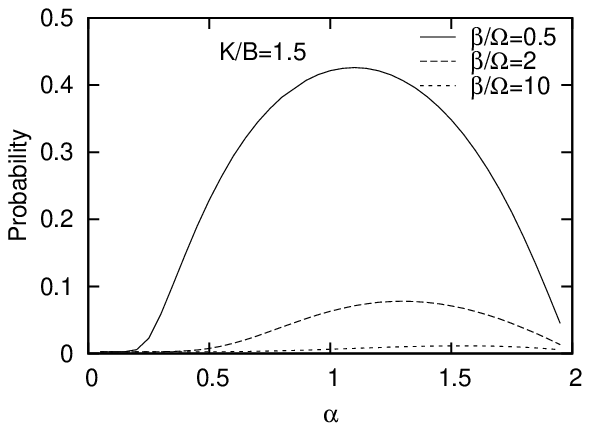} %
\includegraphics[width=7.5cm,clip]{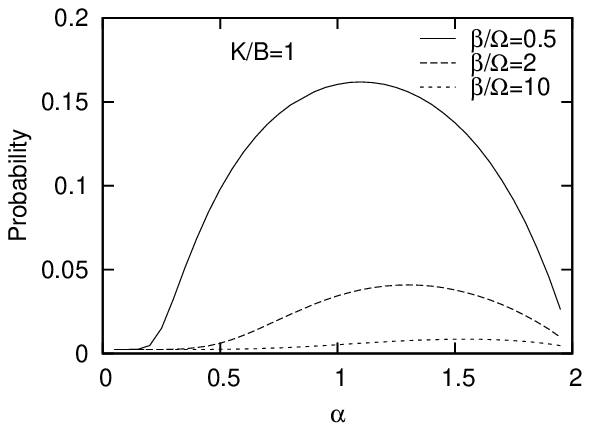} %
\includegraphics[width=7.5cm,clip]{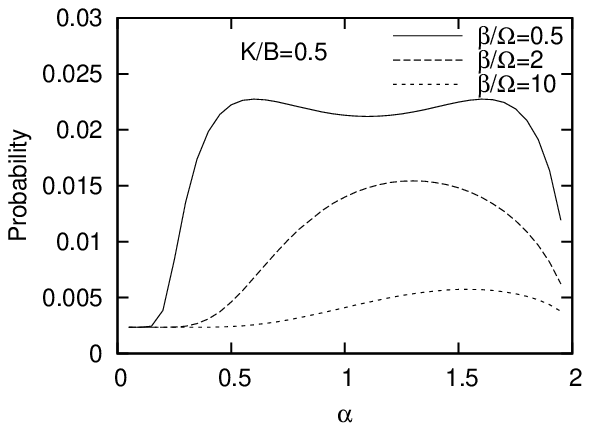} %
\includegraphics[width=7.5cm,clip]{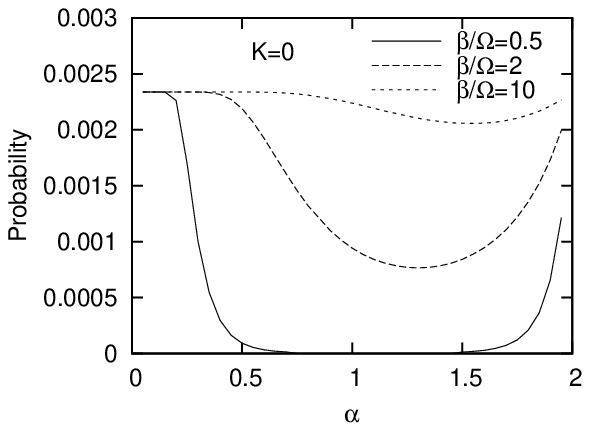}
\caption{The probability Eq. (\ref{prob3}) for the non-Ohmic FMF Eq.
(\ref{nonohm}) is plotted versus the friction $\alpha$. The
temperature is taken so that $T/B=0.25$ and each figure is plotted
with different initial kinetic energy $K/B=1.5,\,1,\,0.5,\,0$ as
labeled on them.} \label{figc2}
\end{figure*}

Our expressions for the over-passing probability are valid for
non-Ohmic dissipation as well, since the characteristic function Eq.
(\ref{D}) again has only one positive root. The Figure \ref{figb2}
shows the plot of the normalized root $y$ of Eq. (\ref{D}) versus
the parameters $\alpha$. The corresponding probabilities are
indicated in Figure \ref{figc2}. The effective friction
$\beta_{\text{eff}}$ is enhanced for the very subdiffusive or very
superdiffusive systems.

We emphasize that our study is limited only to Gaussian
distributions and hence does not include systems exhibiting
non-Gaussian anomalous diffusion like L\'evy flights.

\subsection{Quantum effects}
\begin{figure}[htb]
\includegraphics[scale=0.9,clip]{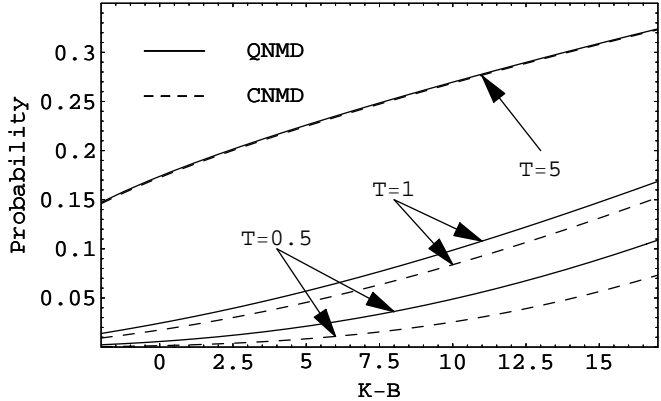}
\caption{The probabilities for the quantum and classical systems are
plotted versus the initial kinetic energy $K$ with respect to the
barrier height $B$ for the temperatures $T= 0.5, 1, 5$. The quantum
non-Markovian and the classical non-Markovian diffusions are
indicated by solid and dashed lines, respectively.}
\label{quaclaprob}
\end{figure}

The previous three sections are dealing with the effects of the FMF
and hence of the normalized root $y$ on the probability. Here, we
investigate the effects of quantum noise on the dynamics. For this
purpose we consider the exponential FMF Eq. (\ref{exp1}) with the
corresponding root Eq. (\ref{b}). The over-passing probability is
obtained by substituting this root and the Fourier transform of the
FMF into Eq. (\ref{quanprob}). In order to compare our results with
some previous studies we consider the fusion reaction of $^{48}Ca$
and $^{238}U$ nuclei with the same parameter set
\cite{Takigawa,Ayik,Washiyama}. The friction coefficient is taken as
$\beta/\Omega= 3.29$, the memory time is $\Omega\tau=1/15$, the
curvature parameter of the conditional saddle is $\Omega=1$ and the
barrier height with respect to the initial position is $B=4$, in
arbitrary units. The comparison of the probabilities for the
classical and quantum systems is shown in Figure \ref{quaclaprob}
which is in good agreement with the previous studies
\cite{Takigawa,Ayik,Washiyama}. At low temperatures, the
over-passing probability is enhanced since the variance of the
position is larger when the quantum effects are included. At high
temperatures, the classical over-passing probability is recovered.

\section{Conclusion}

The probability of a system to diffuse over a barrier is an
important quantity in many research subjects such as activation
processes in chemical physics, fusion and fission reactions as well
as giant resonances in nuclear physics. In this work, we consider
the evolution of a single-relevant variable according to the
classical and quantal GLE. When the potential barrier has the shape
of an inverted parabola, the asymptotic value of the over-passing
probability is given by the complimentary error function according
to Eq. (\ref{prob2}). We show that in the case of classical GLE the
asymptotic value of the over-passing probability is determined by a
single dominant root $y=s_1/\Omega$ of the "characteristic function"
$D(s_1)=0$, and given by a  simple expression Eq. (\ref{prob3}). The
details of dissipation mechanism and memory effects enter into the
expression only through the dominant root of the characteristic
equation. One of the results we found is that for the initial
kinetic energies $K$ less than the barrier height $B$, which is the
case for many physical situations, the over-passing probability of
the diffusion due to the thermal fluctuations becomes maximum when
the dominant root fits the condition given by Eq. (\ref{y_max}).
This is a result of the competition between dissipation and
fluctuation. In the case of quantal GLE, the asymptotic value of the
over-passing probability has the same structure as the classical
one, except it involves a quantity which is determined by a
numerical integration over the spectral density. The probability is
enhanced at low temperatures where the quantum effects are relevant.
The expression for the over-passing probability, Eq. (\ref{prob3})
in the classical limit and Eq. (\ref{quanprob}) in quantal
framework, are valid for a general FMF provided that the FMF has a
well defined Laplace transform.

It is shown that the oscillatory behavior of the FMF can have an
important impact on the factor $y$ and hence on the over-passing
probability. For oscillation frequencies $\lambda$ less than the
critical value Eq. (\ref{lambda}), the non-Markovian dissipation
$\beta_{\text{eff}}$ is reduced with respect to the Markovian one
$\beta$. Whereas for frequencies exceeding the the critical value,
the non-Markovian dissipation is enhanced.

Our formulation also covers systems exhibiting anomalous diffusion
with Gaussian noises. In this case, the static friction coefficient
$\beta^{(0)}$ is zero or infinity for superdiffusive or subdiffusive
systems, respectively. The feature allows to distinguish easily
between these anomalies. The effective friction $\beta_{\text{eff}}$
is enhanced for subdiffusive systems $\alpha<1$ with respect to the
Markovian friction coefficient $\beta$. Whereas there is a minimum
of the effective friction or similarly maximum of the non-Ohmic root
$y^{(\text{NO})}_{\text{max}}$  which changes with $\beta/\Omega$ in
the superdiffusive region $\alpha>1$.

\begin{acknowledgements} S.A. gratefully acknowledges the Physics
Department of Middle East Technical University, D.B. acknowledges
RCNP of Osaka University, and B.Y. acknowledges GANIL, RCNP of Osaka
University and Physics Department of Tohoku University for the warm
hospitality extended to them during their visits. The authors also
thank Noboru Takigawa, Denis Lacroix and Kouhei Washiyama for
fruitful discussions. This work is supported in part by the US DOE
Grant No. DE-FG05-89ER40530, in part by the BDP grant from "Turkish
Science and Technology Council" (TUBITAK), and in part by the JSPS
Core-to-Core Program for Exotic Femto Systems and by JSPS Grant No.
18540268.
\end{acknowledgements}

\end{document}